\documentclass[twocolumn,showpacs,preprintnumbers,amsmath,amssymb]{revtex4}

\usepackage{graphicx}

\begin{document}


\title{External-Feedback Laser Cooling of Gases}

\author{Vladan Vuleti\'{c}, Adam T. Black, and James K. Thompson}
\affiliation{Department of Physics, MIT-Harvard Center for Ultracold
Atoms, and Research Laboratory of Electronics, Massachusetts
Institute of Technology, Cambridge, Massachusetts 02139, USA}

\date{\today}

\begin{abstract}
We analyze the laser cooling of polarizable particles by continuous
dispersive position detection and active feedback. The
magnitude of the dissipative force is proportional to the particles' photon scattering rate into the detector, while its velocity dependence is
determined by the programmable frequency dependence of the loop
gain. The method combines final temperatures near the recoil limit
with large velocity capture range, and is applicable to multilevel
atoms or molecules.
\end{abstract}

\maketitle

Low-temperature, high-brightness sources of particles are of
interest in many areas of physics. In particle accelerators,
stochastic cooling has been a key technology to increase collision
rates \cite{vanderMeer85}. In a single-particle picture, an
electrical signal that depends on a particle's motion generated in
one part of the accelerator ring is used to apply a correction force
further downstream. In view of Liouville's theorem, which forbids
compression of phase space volume by conservative forces, the
cooling of a finite-number sample has been explained as an expulsion
of the empty phase space between occupied points
\cite{vanderMeer85}.  Feedback cooling has been applied to other
systems including a resonant gravity gradiometer \cite{Forward78}, a
single mirror vibration mode via radiation pressure
\cite{Mancini98,Cohadon99}, and a single axially oscillating
electron in a Penning trap \cite{DUrso03}.  Feedback concepts have
been further extended to the quantum domain
\cite{Wiseman94,Wiseman94a,Mancini98,Wallentowitz02,Ivanov03}.  In
order to increase the cooling bandwidth, an optical version of
stochastic cooling has been proposed for accelerators
\cite{Mikhailichenko93}.

For atomic samples, Raizen {\it et al.} have proposed to optically
measure fluctuations of the center-of-mass momentum, and to apply
momentum kicks with laser-induced dipole forces \cite{Raizen98}.
Balykin and Letokhov have suggested a cooling scheme based on
velocity measurement and separation for individual atoms
\cite{Balykin01}. Recently, Steck {\it et al.} have analyzed quantum
feedback cooling of a single atom strongly coupled to a resonator
\cite{Steck04}. In cavity experiments, single atoms have been
trapped by applying feedback \cite{McKeever03,Fischer02}, while in
an optical lattice, a sample's center-of-mass oscillation has been
damped \cite{Morrow02}.

In this Letter, we analyze the laser cooling of a gas of polarizable
particles by continuous dispersive position measurement and optical
feedback. In particular, {\it the sign and velocity dependence of
the dissipative force are determined exclusively by the sign and
frequency dependence of the loop gain}. The force is proportional to
the radiation pressure associated with photon scattering into the
detector, but is otherwise independent of the target level
structure. We derive closed-loop heating rates due to light and
photodetector quantum noise, and due to the thermal motion of the
sample. Using an optical resonator for signal enhancement in
combination with nearly shot-noise-limited photodetection, a large
variety of atomic or molecular samples can be cooled to temperatures
near the photon recoil energy.

In the following we adopt the single-particle viewpoint of
stochastic cooling \cite{vanderMeer85}, where the signal generated
by a single atom is propagated through the linear feedback loop, and
the thermal motion of the other atoms in the sample constitutes a
source of heating. Consider an atom of mass $m$ moving as
$x_\mathrm{a}$=$vt$ at constant velocity $v$ in a weak periodic
potential $V(x,t$)=$U(t)\cos2kx$ whose depth $U(t) \ll mv^2/2$ is
influenced by the atom's motion. Since the time variation of the
force on the atom $f(t)$=$2kU(t)\sin2kvt$ arises both from the
spatial variation $\cos2kx$ of the potential, and from the
dependence of the depth $U(t)$ on the atom's trajectory, a frequency
component of $U(t)$ in phase with the atomic-motion-induced force
variation $\sin2kvt$ produces a non-zero average force that can heat
or cool the atom. To establish this type of dissipative system, we
propose to use a standing light wave to both monitor the position
$x_\mathrm{a}(t)$ of the polarizable atom by means of its index of
refraction \cite{Horak97,Vuletic00,Fischer02}, and to generate a
periodic potential of adjustable depth $U(t)$ via the light shift.

\begin{figure}
\includegraphics[width=3in]{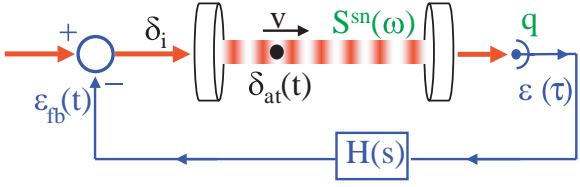}
\caption{Schematic of continuous-feedback cooling. The moving atom's
effective index of refraction periodically detunes the standing-wave
resonator by $\delta_\mathrm{at}(t)$.  The fractional change in
cavity transmission $\varepsilon(t)$ acts as an error signal that is applied to change
the input power by a fraction $\varepsilon_\mathrm{fb}(t)$.  The
feedback open-loop gain $H(s)$ is set by an external electronic
circuit. The input light is initially detuned by $\delta_\mathrm{i}$
from cavity resonance.} \label{setup}
\end{figure}

Viewed in the time domain, the atom is cooled because it finds
itself moving uphill on a deeper potential than experienced when
moving downhill (see also Refs. \cite{Horak97,Vuletic00}).  Of
course, this is only true for an appropriately chosen transfer
function mapping $x_\mathrm{a}(t)$ onto $U(t)$.  In the frequency
domain picture, both the moving atom and the feedback actuator
modulate the light wave at the Doppler frequency $2kv$. Under
appropriate conditions, amplitude and phase modulation from both
processes interfere to produce asymmetric sidebands in the light
exiting the system. If the blue Doppler sideband is stronger, the
atom will be cooled at the rate at which energy is carried away by
the frequency-shifted light. The atom-light interaction, and
consequently the cooling force, can be increased by means of an
optical resonator (Fig. \ref{setup}).

The resonator of finesse $F$ and field decay rate constant
$\gamma_\mathrm{c}$ supports a standing-wave Gaussian $TEM_{00}$
mode of wave number $k$ with waist $w$. Incident light produces an
electric field of amplitude $2E_\mathrm{c}$ at an antinode ($x$=$0$)
inside the resonator. An atom with complex polarizability $\alpha$
moving on the resonator axis coherently scatters photons into free
space at an average rate $\Gamma_\mathrm{sc}$=$k^3 \vert Re(\alpha)
E_\mathrm{c} \vert^2 /(6\pi \epsilon_0 \hbar)$, and experiences an
optical potential $U_0 \cos 2kx$ of depth $U_0$=$-\vert E_\mathrm{c}
\vert^2 Re(\alpha)/2 $. Conversely, the forward-scattered radiation
by the spatially varying induced atomic dipole, or equivalently, the
atom's index of refraction, results in an atomic-position-dependent
detuning $\delta_\mathrm{at}$=$ \zeta \gamma_\mathrm{c}
\cos2kx_\mathrm{a}$ of the resonator from its average resonance
frequency $\omega_\mathrm{c}$ \cite{Horak97,Vuletic00}. Here
$\zeta$=$ \hbar \eta \Gamma_\mathrm{sc}/U_0$ is a dimensionless
parameter characterizing the atom-cavity coupling, and the quantity
$\eta$=$6F/(\pi k^2 w^2)$ can be interpreted as the fraction of
photons scattered into one direction of the resonant cavity.

For incident light detuned by an amount $\delta_\mathrm{i}$=$ck -
\omega_\mathrm{c}$ relative to the resonator, the time-varying
detuning $\delta_\mathrm{at}(t)$ of the resonator by the moving atom
changes the resonator transmission by a fraction $\varepsilon(t)$.
The transmitted power is measured and used as an error signal in a
feedback loop to adjust the incident power by a fraction
$-\varepsilon_\mathrm{fb}(t)$ (Fig. \ref{setup}). In order to ignore
cavity-induced forces \cite{Horak97,Vuletic00}, we take the
resonator linewidth $2\gamma_\mathrm{c}$ to be much larger than both
the feedback bandwidth and the Doppler frequency $2kv$, such that
the intracavity power $P_\mathrm{c}(t)$ adjusts instantaneously to a
value determined by the total light-resonator detuning
$\delta_\mathrm{t}(t)$=$\delta_\mathrm{i}-\delta_\mathrm{at}(t)$.
Then the fractional deviation
$\varepsilon(t)$=$P_\mathrm{c}(t)/P_0-1$ of the intracavity power
from its unperturbed value $P_0$ (for
$\delta_\mathrm{at}$=$\varepsilon_\mathrm{fb}$=$0$) is given by
\begin{equation}
\varepsilon(t) = \frac{\gamma_\mathrm{c}^2+
\delta_\mathrm{i}^2}{\gamma_\mathrm{c}^2+ \delta_\mathrm{t}^2(t)}
\bigl(1- \varepsilon_\mathrm{fb}(t) \bigr)-1.
\end{equation}
Assuming $\vert \delta_\mathrm{at} \vert /\gamma_\mathrm{c},
\varepsilon_\mathrm{fb} \ll 1$, we can linearize the fractional
change $\varepsilon(t)$ in resonator transmission,
\begin{equation}
\varepsilon(t) = r \gamma_\mathrm{c}^{-1} \delta_\mathrm{at}(t) -
\varepsilon_\mathrm{fb}(t).
\end{equation}
In this approximation, the moving atom modulates the intracavity
power by an amount $r \delta_\mathrm{at}/\gamma_\mathrm{c}$
proportional to the normalized resonator slope
$r$=$2\delta_\mathrm{i} \gamma_\mathrm{c}
/(\gamma_\mathrm{c}^2+\delta_\mathrm{i}^2)$, while the feedback loop
adjusts the incident power by a fraction
$-\varepsilon_\mathrm{fb}(t)$. If the atom's kinetic energy far
exceeds the light shift $U_0$, then to lowest order
$\delta_\mathrm{at}(t)$ is determined by the atom's unperturbed
motion $x_\mathrm{a}$=$vt$. We introduce the open-loop feedback gain
$H(s)$ for the Laplace transformed quantities
$\tilde{\varepsilon}_\mathrm{fb}(s), \tilde{\varepsilon}(s)$ via
$\tilde{\varepsilon}_\mathrm{fb}$=$H \tilde{\varepsilon}$. Then
$H(i\omega)$=$H_1(\omega)+iH_2(\omega)$, with real and imaginary
parts $H_1$ and $H_2$, respectively, is the complex gain in the
frequency domain. If the loop is stable, the steady-state solution
in the time domain is given by
\begin{equation}
\varepsilon(t) = r \zeta \frac{\bigl( 1+H_1(2kv) \bigr) \cos 2kvt +
H_2(2kv) \sin 2kvt}{\vert 1+H(2ikv) \vert^2}. \label{epst}
\end{equation}
$H_1(2kv)$ and $H_2(2kv)$ are the open-loop gain in phase and in
quadrature with the atomic-motion-induced intensity modulation
$\delta_\mathrm{at}(t)$=$ \zeta \gamma_\mathrm{c} \cos2kvt$,
respectively. They determine the closed-loop signal
$\varepsilon(t)$, and thereby the time variation of the
optical-potential depth $U(t)$=$U_0(1+\varepsilon(t))$. In the limit
$U_0 \ll mv^2/2$, work is done on the atom at a rate
$\dot{W}$=$\varepsilon(t)f_\mathrm{u}(t)v$ to lowest order, where
$f_\mathrm{u}=2k U_0 \sin2kvt$ is the unperturbed force. The
component $\sin2kvt$ of $\varepsilon(t)$ in phase with
$f_\mathrm{u}$ produces a dissipative force, whose spatial average
$f$=$\langle \dot{W} \rangle /v$ can be written as
\begin{equation}
f(v) = \hbar k \eta \Gamma_\mathrm{sc} \frac{r H_2(2kv)}{\vert
1+H(2ikv) \vert^2}. \label{fcool}
\end{equation}
This expression, derived without invoking the two-level or
rotating-wave approximations, is valid for arbitrary laser detuning
from atomic resonances below saturation. The friction force $f$
depends on target parameters exclusively through the scattering rate
$\Gamma_\mathrm{sc}$ and is controlled by the average intracavity
intensity. $f$ is proportional to the rate of photon momentum
transfer $2\hbar k \eta \Gamma_\mathrm{sc}$ due to backward
scattering into the resonator at rate $\eta \Gamma_\mathrm{sc}$,
multiplied by a dimensionless function of the atomic velocity.  In
particular, the sign and velocity dependence of $f$ are determined
by the frequency-dependent loop gain $H(i\omega)$. They are
independent of the atom's level structure, such that atoms in
different internal states, or different species, can be cooled
simultaneously. Cooling occurs if the product of the resonator line
slope $r$ and the quadrature loop gain $H_2(2kv)$ at the Doppler
frequency $2kv$ is negative. The force is maximized at a
resonator-light detuning $\delta_\mathrm{i}$=$\pm\gamma_\mathrm{c}$
that gives the largest slope $r$=$\pm 1$. In the following we assume
$\delta_\mathrm{i}$=$-\gamma_\mathrm{c}$ ($r$=$-1$), such that
$H(s)>0$ corresponds to negative feedback.

The cooling force $f$ is proportional to the quadrature component
$\sin2kvt$ of the intracavity light modulation in closed loop, given
by $H_2/\vert 1+H \vert^2$. In particular, for very small or very
large open-loop gain ($\vert H \vert^2$=$H_1^2+H_2^2 \ll 1$ or
$\vert H \vert^2 \gg 1$), this relevant quadrature of the intensity
variation, and hence $f$, will be small. The velocity-dependent term
takes on its maximum value $1/(2+2H_1)$ when the quadrature gain
$H_2$ and the in-phase gain $H_1$ are related by $\vert H_2
\vert$=$\vert 1+H_1 \vert$. When $\vert 1+H_1 \vert \ll 1$, the
feedback loop regeneratively amplifies the intensity variation
caused by the moving atom, so that the force is increased by a
factor $\vert 1+H_1 \vert^{-1} \gg 1$ compared to the case of small
in-phase gain $\vert H_1 \vert \ll 1$. However, the heating of the
atom due to noise amplification is then also increased, as we show
below.

\begin{figure}
\includegraphics[width=3in]{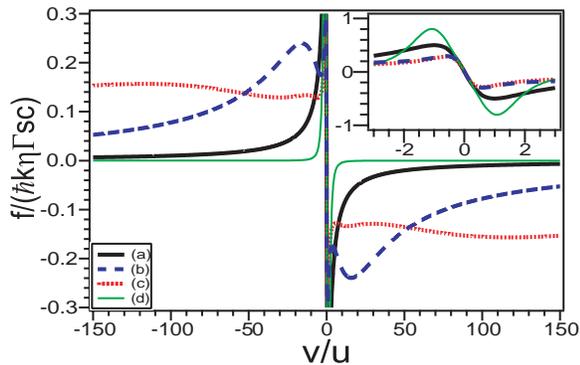}
\caption{Feedback cooling force versus atomic velocity for various
loop gains. The feedback loop gain can be chosen (as in a,
b, or c) to produce a much larger velocity capture range than
conventional Doppler cooling (d), while still yielding the same
friction coefficient near zero velocity (see inset).  The open-loop
gains plotted are (a) $H(s)$=$s$ (differentiator), (b)
$H(s)$=$s(1+s/10)/(1+8s/10)$, (c)
$H(s)$=$s(1+s/10)(1+s/100)/((1+8s/10)(1+s/20))$, and (d)
$H(s)=s(1+s/2)$ reproducing the Doppler force. Here $s$=$iv/u$,
where $u$ sets the unity gain frequency for the differentiator loop.}
\label{force}
\end{figure}

The simplest stable cooling loop is a differentiator
\cite{Fischer02}, $H_\mathrm{d}(i\omega)$=$i \omega/(2ku)$, with
unity gain frequency $2ku$. (The gain can be rolled off at high
frequency outside the velocity range of interest to the cooling.) We
further assume for simplicity that $\Gamma_\mathrm{sc}$ is made
independent of atomic velocity by detuning the laser frequency from atomic
resonances by an amount much greater than the Doppler shift $2kv$.
The differentiator-induced cooling force $f_\mathrm{d}$ is then
given by
\begin{equation}
f_\mathrm{d}(v) = - \hbar k \eta \Gamma_\mathrm{sc} \frac{u
v}{u^2+v^2}, \label{diff}
\end{equation}
The maximum force $\hbar k \eta \Gamma_\mathrm{sc}$, attained for
the unity-gain velocity $v$=$u$, is the same as in Doppler cooling
at the photon scattering rate $\eta \Gamma_\mathrm{sc}$, but the
differentiator force, falling off as $1/v$ rather than $1/v^3$,
provides a substantially larger velocity capture range than the
Doppler force. Fig. \ref{force} also shows that higher-order loops
can substantially further increase the velocity capture range while
maintaining the low-velocity friction coefficient
$\alpha_0$=$\partial f / \partial v$ that determines the final
temperature. $\alpha_0$ can be increased in real time simply by
increasing the loop gain $H$, i.e. by reducing $u$. We also note
that a loop $H(s) $=$(s / \Gamma')(1+ (s/ (2 \Gamma'))$ reproduces
the velocity dependence of the Doppler force \cite{Gordon80} for a
(fictional) transition with linewidth $2 \Gamma' < 2
\gamma_\mathrm{c}$.

Technical or quantum noise in the light-induced potential $U(t)$
will heat an atom moving at velocity $v$ in proportion to the
spectral noise density at the modulation frequency $2kv$ of the
unperturbed force. We define the single-sided fractional spectral
noise density $S_\mathrm{c}(\omega)$=$(2/\pi) \int_0^\infty dt
\langle \hat{\varepsilon}(0) \hat{\varepsilon}(t) \rangle \cos\omega
t$, normalized such that $\int_0^\infty d\omega
S_\mathrm{c}(\omega)$=$\langle \hat{\varepsilon}^2(0) \rangle$ is
the mean-square fractional intensity noise, and calculate the
heating power due to momentum diffusion \cite{Gordon80},
\begin{equation}
\dot{W}_{fluct} = \frac{\pi k^2 }{m} U_0^2 S_\mathrm{c}(2kv).
\label{heating}
\end{equation}
This general formula describes the heating of an atom moving at
velocity $v$ due to classical intracavity intensity noise. For
quantum noise, Eq. (\ref{heating}) needs to be multiplied by a
factor of two to take into account fluctuations of the induced
atomic dipole \cite{Gordon80}. Assuming negligible technical noise,
the shot noise of the light incident on the cavity with power
$P_\mathrm{i}$ corresponds to a fractional power spectral density
given by $S_\mathrm{i}^\mathrm{psn}$=$\hbar ck /(P_\mathrm{i} \pi)$.
In closed loop, the intracavity noise density is
$S_\mathrm{c}^\mathrm{psn}(\omega)$=$S_\mathrm{i}^\mathrm{psn} /
\vert 1+H(i \omega) \vert ^{2}$ for $\omega \ll \gamma_\mathrm{c}$.
In addition, shot noise in the detector photocurrent at limited
quantum efficiency $q<1$ arising from the undetected photons will
produce a false error signal, causing the feedback loop to generate
(classical) intracavity intensity noise. This photoelectron
detection noise corresponds to an uncorrelated fractional noise
density of magnitude
$S^\mathrm{esn}$=$S_\mathrm{i}^\mathrm{psn}(q^{-1}-1)$ in open loop,
and $S_\mathrm{c}^\mathrm{esn}$=$S^\mathrm{esn} \vert H \vert^2 /
\vert 1+H \vert^2$ in closed loop.  The total shot noise from these
two sources
$S_\mathrm{c}^\mathrm{sn}$=$S_\mathrm{c}^\mathrm{psn}+S_\mathrm{c}^\mathrm{esn}$,
expressed in terms of the intracavity power
$P_\mathrm{c}$=$P_\mathrm{i} F/(2 \pi)$ for the detuned cavity with
$\vert \delta_\mathrm{i} \vert$=$\gamma_\mathrm{c}$, is given by
\begin{equation}
S_\mathrm{c}^\mathrm{sn}(\omega) =  \frac{F \hbar c k}{2 \pi^2
P_\mathrm{c}} \frac{1 + \vert H(i\omega) \vert^2 (q^{-1}-1)}{\vert
1+H(i\omega) \vert^2}. \label{spectraldensitySN}
\end{equation}
Eq. (\ref{heating}) then yields a closed-loop heating rate
$\dot{W}_\mathrm{sn}$ due to shot noise
\begin{equation}
\dot{W}_\mathrm{sn} =  E_\mathrm{r} \eta \Gamma_\mathrm{sc} \frac{2
+ \vert H(2ikv) \vert^2 (q^{-1}-1)}{\vert 1+H(2ikv) \vert^2},
\label{heatingSN}
\end{equation}
where $E_\mathrm{r}$=$\hbar^2k^2/(2m)$ is the recoil energy. We see that $\dot{W}_\mathrm{sn}$ can be
expressed as the recoil heating rate $2 E_\mathrm{r} \eta
\Gamma_\mathrm{sc}$, multiplied by a dimensionless function that
depends on the loop gain $H$ and the detector quantum efficiency
$q$. In the absence of feedback ($H$=$0$), Eq. (\ref{heatingSN})
reproduces the well-known recoil heating \cite{Gordon80} due to
scattering into the resonator, where each backscattering event,
occuring at a rate $(\eta/2) \Gamma_\mathrm{sc}$ for the detuned
cavity, heats the atom by an amount $4E_\mathrm{r}$. For unity
quantum efficiency ($q$=$1$) the feedback loop reduces (increases)
this heating by suppressing (enhancing) the intracavity fluctuations
for $\vert 1+H \vert >1$ ($\vert 1+H \vert <1$). For $q<1$, the
feedback loop also generates intensity noise originating from random
photoelectron detection. In addition to the cavity heating
$\dot{W}_\mathrm{sn}$, scattering into free-space heats the atom at
a rate $\dot{W}_\mathrm{fs}$=$2E_\mathrm{r} \Gamma_\mathrm{sc}$
\cite{Gordon80}.

For the differentiator loop $H_\mathrm{d}(i\omega)= i \omega/(2ku)$,
a unity-gain velocity $u$=$(q^{-1}-1+2\eta^{-1})\hbar k/m$ minimizes
the temperature $T_\mathrm{d}$, yielding
\begin{equation}
k_\mathrm{B} T_\mathrm{d}= 4 E_\mathrm{r} \frac{(1 + \eta)(q^{-1} -
1 + 2\eta^{-1})}{\eta}, \label{Tsingle}
\end{equation}
where $k_\mathrm{B}$ is Boltzmann's constant. Thus for good
atom-cavity coupling, $\eta = 1$, the final temperature is limited
by the finite detection efficiency to $8(1+q^{-1})
E_\mathrm{r}/k_\mathrm{B}$. This implies that except for the
lightest atoms and molecules, microkelvin temperatures can be
reached even for limited detector quantum efficiency $q$.

For a thermal sample consisting of $N$+1 atoms at temperature $T$,
any chosen probe atom will be heated by the intracavity intensity
noise induced by the other $N$ randomly moving atoms. The
corresponding closed-loop spectral noise density
$S_\mathrm{c}^N(\omega)$, expressed in terms of the sample's thermal
velocity $v_\mathrm{th}$=$(k_\mathrm{B} T/m)^{1/2}$, is
\begin{equation}
S_\mathrm{c}^N(\omega) = \frac{N \zeta^2}{\sqrt{8 \pi}
kv_\mathrm{th} } \frac{\exp\bigl(- \omega^2/(8 k^2
v_\mathrm{th}^2)\bigr)}{\vert 1+H(i\omega) \vert^2}, \label{stildeN}
\end{equation}
resulting in a collective heating rate $\dot{W}_\mathrm{N}$ for an
atom with velocity $v$ given by
\begin{equation}
\dot{W}_\mathrm{N}(v) = \frac{E_\mathrm{r} \eta
\Gamma_\mathrm{sc}}{\vert 1+H(2ikv) \vert^2} \frac{\sqrt{2 \pi} N
\eta \Gamma_\mathrm{sc}}{2kv_\mathrm{th}} \exp\Bigl(- \frac{v^2}{2
v_\mathrm{th}^2}\Bigr). \label{heatingN}
\end{equation}
As is typical of stochastic cooling, the heating overwhelms the
cooling at too large atom number $N$ or too large cooling speed,
i.e. photon scattering rate $\Gamma_\mathrm{sc}$. For a
differentiator loop, we find that the net cooling power averaged
over the thermal sample, $\langle \dot{W}_\mathrm{d}
\rangle$=$\langle f_\mathrm{d} v+\dot{W}_\mathrm{N} \rangle$  is
maximized for $u \approx v_\mathrm{th}$. The cooling rate constant,
$\gamma_\mathrm{d}$=$-2\langle \dot{W}_\mathrm{d} \rangle
/(k_\mathrm{B} T)$ can then be written in the form
\begin{equation}
\gamma_\mathrm{d} \approx \frac{kv_\mathrm{th}}{6N}
(2\bar{\Gamma}_\mathrm{cav}-\bar{\Gamma}_\mathrm{cav}^2)
\label{coolrate}
\end{equation}

\noindent where
$\bar{\Gamma}_\mathrm{cav}$=$2N\eta\Gamma_\mathrm{sc}\hbar/(k_\mathrm{B}
T)$ is the total scattering rate into the cavity normalized to the
sample temperature. The cooling rate is maximized for
$\bar{\Gamma}_\mathrm{cav}$=1, yielding a rate constant
$\gamma_\mathrm{d} = k v_\mathrm{th}/(6N)$. We see that the thermal
Doppler broadening $k v_\mathrm{th}$ takes the role of the
stochastic-cooling bandwidth \cite{vanderMeer85}, and that the
cooling of smaller subsamples can proceed faster
\cite{vanderMeer85,Raizen98}. For a sample of $N$=$10^8$
magnetically trapped CaH molecules at 0.4 $K$ \cite{Weinstein98a}
cooled with light of wavelength 760 nm, the optimum cooling rate is
$\gamma_\mathrm{d}$=0.1 s$^{-1}$, attained at a photon scattering
rate $\eta \Gamma_\mathrm{sc}$=$3 \times 10^2$ s$^{-1}$. (Here we
are assuming that the different degrees of freedom are mixed
\cite{vanderMeer85}, e.g., by the trapping potential.) An ensemble
of $10^6$ molecules at room temperature can be cooled at a rate
$\gamma_\mathrm{d}$=340 s$^{-1}$ for a photon scattering rate of
$\eta \Gamma_\mathrm{sc}$=$3 \times 10^7$ s$^{-1}$.  Eq.
\ref{coolrate} implies that the differentiator-induced cooling of a
thermal beam containing $N$ atoms with light of wavelength $\lambda$
requires a characteristic length $L$=$6 N \lambda / \pi$,
independent of the initial velocity of the beam, since the cooling
bandwidth $kv_\mathrm{th}$ is larger for a faster beam.

It may seem surprising that an error signal proportional to the
atom's index of refraction generates a cooling force that is
proportional to the photon scattering rate into the detector. The
reason is that the index of refraction arises from the superposition
of the forward-scattered field with the much larger incident field.
The dispersive signal can thus be viewed as a homodyne method to
measure the small forward-scattered field. Homodyne detection,
however, cannot be used to improve over the shot-noise limit
resulting from direct detection of the scattered photons. The role
of the resonator is to enhance the detection signal-to-noise ratio
\cite{Balykin01}, or equivalently, the photon scattering rate $\eta
\Gamma_\mathrm{sc}$ into the detector. In the absence of the
resonator, $\eta$ is replaced by the detection solid angle $\Delta
\Omega$=$3/(k^2w^2)$, resulting according to Eq. (\ref{Tsingle}) in
much higher temperatures. This may explain why stochastic cooling
has not been observed in free-space experiments using laser beams
with large waist $w \gg k^{-1}$ for detection
\cite{Raizen98,Morrow02}.

In conclusion, we have derived simple analytical expressions for the
cooling and heating of a gas interacting with a laser beam whose
intensity is adjusted in response to the atoms' motion. If a
resonator is used for signal enhancement in combination with
photodetection near the quantum limit, microkelvin temperatures can
be reached. By appropriately designing the frequency-dependent loop
gain, large friction coefficients can be combined with a very large
velocity capture range. The friction force scales with the photon
scattering rate by the target atoms, but is otherwise independent of
the target level structure. Therefore the proposed method represents
a promising route for laser-cooling new atomic or molecular species.

This work was supported in parts by the NSF, the ARO, and the Sloan Foundation.


\end{document}